# Promoting Rural Entrepreneurship through Technology: A Case Study using Productivity Enhancing Technology Experience Kits (PETE-Kits)

Matthew W. Rutherford, Brian E. Whitacre, Levi Captain, Sabit Ekin, Julie Angle, Tom Hensley, and John F. O'Hara

*Abstract—Contribution*: Case study of a rural-focused educational program with two components: 1) introducing high school students and teachers to Smart and Connected Technologies (SCTs) that can be used to solve local problems; 2) engaging the local community in supporting local technology-driven entrepreneurship.

*Background*: Rural communities typically lag behind in terms of participation in the digital economy, and use of technology in general. Yet they often have the most to gain, due to high rates of self-employment and lower private-sector job opportunities.

*Research Questions*: Can a broadly-scoped rural technology education program lead to improvements in 1) student and teacher SCT awareness, 2) SCT skills, 3) aspirations for future SCT use directed toward entrepreneurship and overall community wellbeing?

*Methodology*: Our multidisciplinary team used a mixed-methods approach to engage a rural high school robotics team as well as the local community. Over the course of one year, students took part in hands-on-training with SCTs ("PETE-Kits" and associated curriculum) and brainstormed entrepreneurial projects via ideation events. Community members were involved at the beginning and end of the project, including judging a "shark-tank" style event where student business ideas using SCT were presented.

*Findings*: Results from student pre / post activity assessments suggest that the program was effective at increasing comfort with technology and combining technical skills with entrepreneurial opportunities. Post surveys from community members, including teachers, demonstrated clear support for the program and an appreciation of how SCTs / digital skills could benefit the local economy and wellbeing.

## I. INTRODUCTION

LACK of participation in the digital economy is an impediment to societal well-being and production [1], which asymmetrically affects rural communities and is an important component of the "digital divide" [2-4]. Recent U.S. policy efforts have provided significant funding (over $60 billion) to improve rural broadband infrastructure [5] and offer digital skills training to help residents use it productively [6]. Naturally, then, programs designed to improve rural citizens' participation in the digital economy are essential to such policy efforts and to ensure digital career equities. In response to this need, this paper puts forward a rural education program infused with elements from both entrepreneurship and smart and connected technologies (SCT). Through our novel curriculum, we set out to educate, motivate, and enroll key rural stakeholders into this new economy.

In partnership with the rural community of Frederick, OK, our broad goals for the program were three-fold. First, we wanted to witness an improvement in community technology literacy. Second, we desired to improve confidence toward either entrepreneurship or remote work within the entire community. Third, we hoped that the program would demonstrate the ability to be scaled. Overall, we feel that the project was a qualified success, and our results demonstrate preliminary support for the program – an encouraging sign as states explore ways to spend soon-to-be-allocated federal digital equity funds [5]. Our work allows us to reveal new and novel insights about teaching technology literacy and entrepreneurship to rural communities. First, we present our research questions and general approach. Second, we describe our low-cost technology compilation—the PETE-Kit—and its accompanying curriculum and explain how they conceptually

This work was supported in part by the National Science Foundation Smart and Connected Communities Grant #2125393. *(Corresponding author: John O'Hara).*

J. O'Hara is with the School of Electrical and Computer Engineering, Oklahoma State University, Stillwater, OK 74078 USA (email: oharaj@okstate.edu).

L. Captain is with the School of Electrical and Computer Engineering, Oklahoma State University, Stillwater, OK 74078 USA (email: george.captain@okstate.edu).

M. Rutherford is with the Spears School of Business, Oklahoma State University, Stillwater OK 74078 USA (email: matthew.rutherford@okstate.edu).

B. Whitacre is with the Department of Agricultural Economics, Oklahoma State University, Stillwater OK 74078 USA (email: brian.whitacre@okstate.edu).

S. Ekin is with the Department of Engineering Technology & Industrial Distribution, Texas A&M University, College Station TX 77843 USA (email: sabitekin@tamu.edu).

J. Angle is with the School of Teaching, Learning, and Educational Sciences, Oklahoma State University, Stillwater, OK 74078 USA (email: Julie.angle@okstate.edu).

T. Hensley is with Frederick Public High School, Frederick OK 73542 USA (email: thensley@frederickbombers.net)

Supplemental materials can be found at https://osf.io/n5hzb/

Color versions of one or more of the figures in this article are available online at http://ieeexplore.ieee.org

expedite growth in technology literacy. Third, we discuss the results of our research, highlighting our finding of the critical importance of a local and embedded "champion." Finally, we specify the targeted types of entrepreneurial training needed to generate desired results. When combined, these curricular imperatives substantially enhance preparedness of rural individuals' participation in the digital economy. That said, gaps in our knowledge remain and work in this area has likely just begun.

## II. Literature Review

Productivity is foundational to the ability of rural communities to survive and thrive in a competitive technological world economy – but the nature of this productivity is changing [7]. In the past, agriculture and manufacturing were the primary sources of economic productivity in rural communities [8-9]. However, "productivity" is no longer so limited to these sources. Intellectual skills and aspirations are arguably more critical factors in modern productivity [1, 10] and have impact beyond economics and into quality of life and subjective wellbeing. We believe that this is excellent news for rural communities since talent is easily exchangeable with broadband connectivity – through "remote work." That is, rural residents need not relocate to participate in this economy. In this sense, SCTs have the potential to multiply the productive efforts of rural citizens, allowing them to stay in rural areas, have highly successful careers, and improve opportunities for healthcare and civic engagement [11-13]. To this end, whole communities are partnering with tech giants to realize this potential [14].

Unfortunately, though, a mismatch exists between what is presently available and what is needed in terms of human capital in rural areas. This strongly suggests that solutions to rural growth and sustenance issues could be improved through well designed and implemented education or re-education programs focusing on utilizing SCT for entrepreneurial or remote work ventures. While this broad nostrum is apparent, there are many gaps in our knowledge for successful rural technological entrepreneurship education programs.

Fortunately, extant research is clear that technological entrepreneurship skills can be learned [15-16]. This same work also offers insights—however general—into how to devise programs in ways that can impart these skills [17-18]. Here we review this work and then marry it with the unique issues facing rural populations including a lack of federal and state resources for rural public education [19], declining school enrollment and distance to urban areas [20], and difficulties obtaining and implementing new technology [21]. The review will result in a limited set of best practices that will inform our program design and implementation.

### A. Technology Entrepreneurship

The work in technology entrepreneurship education is substantial, including studies emphasizing an engineering approach [22-23], effective methods for "boot camps" [24] and course assessment [25], and the role of human and social capital [26-27]. These studies can provide frameworks for entrepreneurial education that consider the local context – for example, areas with strong social capital can emphasize industry / business partnerships with a broader array of opportunities while those without it may need to focus on cases where clear demand already exists. They also provide distinction between different learning paradigms such as behavioral (typically via lectures or structured exercises) vs. situated (where students are challenged to use critical thinking skills) learning [22, 28].

Even with this prior body of knowledge, large gaps remain. Very little of this extant literature explicitly focuses on rural locations, which come with their own challenges in terms of human / social capital and resources for both public teaching and community engagement. No technology entrepreneurship studies that we are aware of emphasize the community-wide aspect of embracing SCTs for entrepreneurial purposes, or ways to facilitate that relationship.

We follow Bailetti [29], who defines technology entrepreneurship as "an investment in a project that assembles and deploys specialized individuals and heterogeneous assets that are intricately related to advances in scientific and technological knowledge for the purpose of creating and capturing value." The "value-added" element of our educational program hinges on the interplay between the high school students learning about the entrepreneurial potential of SCTs and the local rural community that has a vital role in supporting them. Our curriculum provides touch-points to assess increases in technology literacy across both groups and culminates in a "shark-tank" style event where the students present their ideas to the broader community group.

### B. Rural Entrepreneurship

There is also some existing work in rural entrepreneurship education, but this is far less developed. By definition, rural communities are isolated from urban centers where the majority of resources are focused [30]; and as a result, the literature indicates that rural communities are not as active as their urban counterparts in technology adoption [31-33]. The adoption problem covers the extent to which rural communities have the financial resources, awareness, skills, and aspirations (or collectively, the literacy) to seize the productivity opportunities afforded by smart and connected technologies [1, 34-36].

This 'smart divide' caused by varying levels of adoption has far-reaching implications such as reduced educational and job opportunities, increased food insecurity, and diminished healthcare options. The problem is partially one of availability, where rural infrastructure is lacking because decreased populations discourage the investment of network providers [37-39]. However, simple availability of key technologies (e.g., broadband) does not automatically translate to revitalization or even stability for rural communities. That is, broadband availability is a necessary, but not sufficient condition for bridging this divide [40]. We know that rural citizens are less likely than metropolitan citizens to adopt and use these technologies [41-42] even when these technologies have strong target applications in rural areas. In fact, recent data suggests that adoption rates of both home broadband and smartphones are 7-10 percentage points lower in rural households compared to their urban counterparts [43]. This adoption and use of SCTs by these populations is critical, yet lagging [1, 40, 41].

In sum, both of these literatures (technology and rural entrepreneurship) have shed light on the manifold issues facing rural communities, but there is very little extant knowledge that brings these areas together in a way that might directly inform educational program design. Without a coalescing and simultaneous treatment of rurality, technology, and entrepreneurship; any program or solution will be incomplete. It will be incomplete because all signs point to the fact that technology and entrepreneurship skills must operate in tandem to overcome the productivity challenges in rural settings. An additional sub-theme of this intersection is that of community. The productive convergence of education, entrepreneurship, and technology seems most likely to be successful in locations where the local environment is both understanding and supportive, however this element also lacks adequate study in extant work.

III. RESEARCH QUESTIONS AND APPROACH

This research was intended to determine whether rural communities could be equipped with greater technology literacy and thereby improve their own productivity and wellbeing. Collectively, we defined 'technology literacy' on three legs, as shown in Figure 1: the awareness of SCTs, the skills to use SCTs, and the aspirations to adapt SCTs to entrepreneurship. This also formed the framework of our research questions: to what extent can our PETE-Kits and curriculum 1) *raise rural awareness* of SCTs, 2) *impart or improve skills* to rural citizenry in the use of SCTs, and 3) enhance productivity by *raising the SCT-inspired aspirations* of the rural community. To answer our multifaceted research questions, we endeavored to develop a curriculum that embraces the confluence of technology, entrepreneurship, and rurality with the goal of enhancing participants' knowledge to a point where they can independently and enthusiastically embark upon a career path of entrepreneurship or remote work with the knowledge that the local community is supportive of that effort.

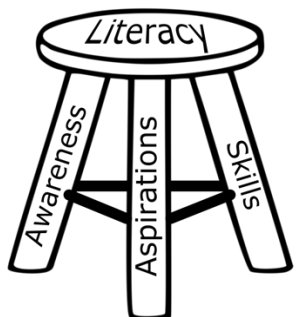

**Fig. 1.** Technology literacy foundation.

With this overarching goal, our SCT-focused curriculum was developed to address two distinct audiences: 1) high school students in a rural community, and 2) the broader population of the community itself. We began by developing specialized teaching tools designated as "Productivity-Enhancing-Technology-Experience Kits" (PETE-Kits). These low-cost technology kits, described below, were designed for hands-on use to quickly raise rural students' and community members' awareness of different types of SCTs and how they might be relevant for life in a rural environment. We partnered with the community to design the PETE-Kits to ensure that they would help students and local residents envision how SCTs relate to their own personal and community-wide ideas, needs, and aspirations.

*A. PETE-Kits and Curriculum Development*

In terms of hardware, PETE-Kits consist of several low-cost technologies such as long-range radios (LoRa), wireless network protocols (WiFi), Bluetooth, microprocessors (e.g., Raspberry Pi or ESP8266), cameras, sensors, actuators, lights, motors, and more. The Supplemental Materials (Supplement A) contains a list of all components (with picture and approximate price) found in each kit. The kits were capable of being tied to the cloud for remote control, data collection, data visualization, monitoring, and automation through services such as Amazon Web Services (AWS). PETE-Kits are meant to enable hands-on exploration, training, and entrepreneurial projects, and thus can take many forms; however, we elected to survey the community's needs, interests, and aspirations before finalizing the list of included components. For example, one individual expressed an interest in monitoring engine emissions, so an air quality sensor was included in each PETE-Kit. PETE-Kits are flexible enough to easily reconfigure and accommodate other communities' unique needs. This modular approach also permits elimination of items of less interest, reducing cost for individuals with lower income.

To begin using the PETE-Kits, our engineering, entrepreneurial, and education team developed an overview of PETE-Kits and their capabilities to educate teachers and students about SCT. This included the following curriculum elements:
- Quick-start guide for how to operate the PETE-Kit and get it connected to a network for the first time
- Glossary of SCT terminology
- Two tutorial projects providing detailed instructions on how to adapt the PETE-Kit for 1) temperature/humidity measurements and 2) recording images with the camera. These tutorials included instructions on how to assemble the PETE-Kit hardware, download and store (or write) software, and operate the PETE-Kit for the application. Extra optional information was provided to explain the operation of the software in more detail.
- Eight 1-page real-world case studies on how technology very similar to the PETE-Kits has been used in entrepreneurial endeavors that profit or improve community wellbeing (for example, remote patient monitoring or soil moisture tracking). Some of these are available in Supplement B.
- PowerPoints slides on the general concept of SCTs and how they are currently being used.
- Ideation protocols for generating business / community entrepreneurial ideas.

- Workshop materials for high school teachers of differing technology skill levels.

### B. Rural Community Selection and Project Methodology

Our multidisciplinary team included faculty members in electrical engineering, entrepreneurship, agricultural economics, and education. As part of a land-grant university, we used our cooperative extension presence in each county of the state to find rural communities that might be a good fit for our project. We sought out a location that was struggling economically (i.e., with population loss and industrial decline) but where high school students had some prior knowledge of technology and where the community would be supportive of our effort. We homed in on rural communities with high school robotics programs and had preliminary discussions with the advisors of those programs, various city representatives (mayor, city manager), and individuals with local resources that might be interested in our work (career tech / community colleges, military bases, telecom providers, known entrepreneurs). A recurring theme as we discussed the project with these communities was, "*These [SCTs] sound promising, but we don't really know what they are, how to use them, or what can be done with them.*"

The community ultimately chosen (Frederick, OK; 2020 population ~3,500) was selected for two main reasons: 1) it fits the economic criteria above; and 2) the eagerness of the participating school (principal and robotics advisor) and clear support for the project from local leaders and stakeholders (mayor, nearby Air Force base personnel, career tech). Frederick has lost roughly 15% of its population since 2010, has a poverty rate of over 25% (OK average 15.6%), and suffered the closure of its only hospital in 2016. Data from the Bureaus of Economic Analysis show significant reductions in employment in retail, health care, and state and local government during the 2010-2020 period. Nonetheless, the community has doubled down on its public education system and established a high school robotics program in 2015. With a passionate advisor/coach, the robotics team has since captured numerous state titles and qualified for world competitions. This "local champion" proved to be critical to the success of our work. Local leaders were also very receptive to the project's inclusion of both the students and broader community.

Our chosen approach was to partner with Frederick and nearby resources to pilot the use of the PETE-Kits described above and then assess the program's impact as laid out in our three formal research questions noted above (raising rural awareness of SCTs, imparting SCT skills on rural students, and raising SCT-inspired aspirations). The research team developed a timeline of events, with student project work sandwiched by community-wide meetings (Figure 2). Data was collected via surveys constructed either for students or general community members (or both) throughout the project.

The initial community kick-off meeting served as an introduction to our program and was largely geared towards showcasing preliminary versions of the PETE-Kits and soliciting buy-in from a wide array of community members. At the end of the initial event, the technology literacy pre-survey (Supplement C) was collected from both students and community members. Its purpose was to assess the extent to which parties: i) already felt skilled in the use of SCTs, ii) gauged their own awareness of SCTs, iii) became enthusiastic about using SCTs for entrepreneurial endeavors and using the PETE-Kit, iv) were able to ideate an entrepreneurial use case involving the PETE-Kit.

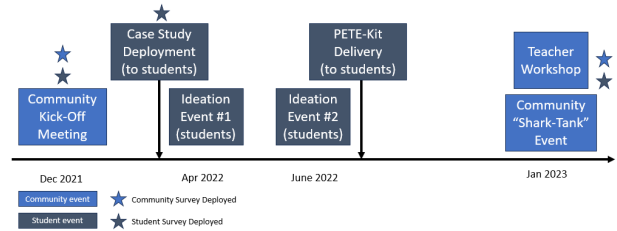

**Fig. 2.** Project event timeline.

Next, we delivered case study documents (Supplement B) to the students (and their advisor/robotics coach) along with a technology literacy assessment (Supplement D). The eight case studies were intended to illustrate how real-world ventures have been built on the foundation of SCTs that are similar to PETE-Kits. Some case study examples included remote oil well monitoring, remote soil moisture monitoring, improved elder care, and in-home patient care or telemedicine. The students were first tasked with reviewing and discussing the eight case studies among themselves and their advisor/robotics coach. For the assessment, the students were tasked with writing an explanatory paragraph, including a specific set of technology-based words, explaining their perception of a previously unseen and hypothetical liquid processing system that is controlled and monitored through the cloud and a local area network tied to a microprocessor. The purpose of this assessment was to gauge the effectiveness of the case studies in improving the awareness of SCTs and the skills to use them. The case study assessment was performed twice during the project: once before PETE-Kits were deployed, and once after.

We then moved on to an in-person ideation event where the goal was to help students generate ideas where the PETE-Kits could be used to help solve a local problem. We followed up with a second ideation event 2 months later, which culminated in teams of students selecting project topics they hoped to work on during the remainder of the year. This event included a more fundamental business model bootcamp, where students were taught the basics of how to turn an idea into a business. At the end of the second event, students had constructed 4 distinct teams of 4-8 members. Each team had a general idea of both the local problem they wanted to use the PETE-Kit to solve and the work that needed to be done to construct a basic business plan.

Following these two student-oriented events, the final PETE-Kits were delivered. Students then had approximately six months to engage with the PETE-Kit and further research their topic, including considering the work required to convert a PETE-Kit into a "Minimum Viable Product" [44] associated with their idea, how their product would differ from existing competitors, and the best way to present their topic to the local community and/or potential funders. This work was student-driven (with advisor/coach oversight) and took place outside of regular school time. Notably, the advisor was well-versed in the associated technologies used, but had minimal expertise in the entrepreneurial angle associated with the project. Nonetheless, the availability of this local champion who was highly respected by the participating students was vital for achieving student buy-in.

The culmination of our program was a "Shark-Tank" styled event where students presented entrepreneurial ideas based on their business and PETE-Kit training to a diverse panel of judges from within and around the rural community. This event was partnered with a 4-hour workshop developed for K-12 teachers in various Oklahoma counties to learn about introducing PETE-Kits in their own classrooms. Teachers attended the workshop in the morning and were then joined in the afternoon by local community members for the student presentations. The student-led teams each gave 15-minute presentations summarizing their chosen SCT application and winners were awarded certificates for various categories, including a "Community Choice" award. Three surveys were administered during the "Shark-Tank" event to all attendees; one was administered both before and after the event, while the other only afterwards (Supplements E & F, respectively).

Our subsequent analysis will examine 1) the effectiveness of our curriculum in conveying desired knowledge 2) the effectiveness of our curriculum in causing desired outcomes. All interactions where data was gathered were reviewed and approved by Oklahoma State University's Institutional Review Board.

## IV. Results

### A. Community Kick-Off Survey Results

The participants at this first event included members of the general community. In all, a total of approximately 48 individuals attended this event which included the Oklahoma State University (OSU) team, high school students, teachers, the mayor, and representatives from local businesses. The purpose of the event was to provide an overview of the planned research project, introduce and showcase the PETE-Kits, and administer a "pre-survey" (Supplement C). After the OSU team introduced the PETE-Kit concept and several explanatory presentations, the pre-survey (Supplement E) was completed by 23 community attendees and attempted to measure (via a 6-point Likert scale, 1 through 6, with 6= "strongly agree") the attendee's opinion on the value of technology literacy and assess their current level of technology skills. Table 1 highlights the two main takeaways: (1) the event was well-received in terms of positioning technology as a benefactor for rural communities (i.e. all scores near or above 5.0), and (2) attendees considered themselves quite aware of SCTs and even generally skilled in their use (i.e. scores around 5.0), but were much less confident (below neutral) in terms of their detailed skills and ideas for how to use the PETE-Kit or SCTs (i.e. scores closer to 4.0 or below). This gave credence to our general hypothesis that rural communities see value in this type of a program but lack the tools to begin on their own. It also suggested that the community perception of technology "awareness" may not have translated to a solid understanding of the purpose of SCT technologies.

### B. Technology Literacy Assessment Results

A total of 39 students participated in the technology literacy assessment (Supplement D) using the case studies. The first 28 students participated before they received PETE-Kits while 11 students participated after the class (as a whole) experienced the PETE-Kits to some extent. Only 2 students participated in both assessments. In addition, numerous students were added to and subtracted from the team during the year-long project. Therefore, the data from both surveys should be viewed as a *starting* assessment of technology literacy, rather than a before-after impact of the PETE-Kits. The assessments were scored by the OSU team using a rubric of eight questions regarding how well the students could identify the nature of the unknown technology system. Some examples are:

1. Did the student identify connectivity between devices?
2. Did the student identify both control and monitoring system functionalities?
3. Was the student able to identify a possible purpose of the system?
4. Did the student use the words from the technology list properly?

TABLE I
STUDENT AND ADULT AVERAGE LIKERT SCORES (1-6) FOR SELECT QUESTIONS AT INITIAL COMMUNITY MEETING EVENT

| Question | Student average (n=12) | Adult average (n=10) |
|---|---|---|
| I have a better understanding of the value of technology with respect to economic aspects of rural life than I did before this event. | 4.83 | 5.70 |
| I have a better understanding of the value of technology with respect to health and the quality of life than I did before this event. | 5.08 | 5.50 |
| Attending this workshop raised my awareness about the types of technology that are available | 5.50 | 5.60 |
| I think achieving technology literacy will support the viability of my rural community | 5.08 | 5.70 |
| I am excited to learn about new technologies using PETE-Kits | 5.42 | 5.30 |
| I have seen good examples of technology-enabling jobs or products in my community | 4.58 | 4.00 |
| I have a strong awareness of modern technology and how it may be leveraged for my benefit | 5.25 | 4.90 |
| I feel skilled at the use of electronic, software, and cloud (Internet) technologies | 5.08 | 4.30 |
| I currently have the technology skills to design a PETE-Kit | 3.92 | 3.20 |
| I have an idea of what I want my PETE-Kit project to measure | 3.33 | 3.10 |

One participant did not report adult or student status.

If the grader could subjectively affirm the student's understanding with 75% confidence on any question, the question was graded with 1 point and otherwise 0 points. The first 28 students scored with an average of 2.96/8.00, while the following 11 students scored with an average of 3.09/8.00, suggesting no significant change. The two students who participated in both assessments improved their scores, going from 3/8 to 4/8 in one case and 2/8 to 5/8 in the other.

The data elicits two conclusions. First, since there was so much change in the student population between assessments, a score of approximately 3/8 (or 38%) appears to be a common starting level of technology literacy. On the other hand, the fourth assessment question (above), regarding the use of technology words, was scored correctly 71% and 63% of the time, in the first and second tranches respectively. This underscores our contention that people are aware of SCT to a much greater extent than they possess the skills to use SCT for practical purposes. Second, the low scores highlight the need to better organize and flesh out the curriculum surrounding the case studies. While the case studies were well-received by the advisor/coach and make a strong and important tie to the entrepreneurial angle, there was no clear correlation to the assessment nor a formal mechanism to check or improve the students' level of their understanding. Therefore, this is an obvious area of needed improvement.

A group of teachers also participated in the technology literacy assessment after reviewing the case studies. Since this was part of the Teacher Workshop (see Figure 2), we summarize those results in the section below.

*C. Teacher Workshop Results*

The teacher workshop was attended by 15 teachers from various regions of Oklahoma. Each participant was provided a PETE-Kit for future use in their own classrooms and an honorarium upon completing the workshop requirements. The workshop began with presentations discussing the overview and rationale behind the project, introductions to technology literacy and related terminology, a description of the PETE-Kit and the included SCT, and finally an overview of the educational link between our curriculum and the Next Generation Science Standards [45]. Teachers then broke out into working groups to review and discuss the eight case studies described above and to participate in the associated technology literacy assessment. The assessment was the same assignment that the students took. Next, during a working lunch, the OSU team led discussions regarding the implementation of the PETE-Kit curriculum in the public-school classroom, and how entrepreneurship education is related and might be included.

The technology literacy assessment had remarkably similar results to those produced by the students. Teachers' average assessment was 2.80/8.00, suggesting an overall lower literacy than the students in regard to SCT. However, at the same time, the teachers were assessed as more adept at using technology-based words in their description of an unknown system. Quantitatively, teachers scored an average of 80% on the fourth assessment question, suggesting the teachers' awareness of SCT was slightly better than students', or that they were simply more skilled in communicating. In connection with the student assessments, we may conclude that technology literacy challenges are very similar across generational gaps, which suggests the PETE-Kit and its curriculum could be productively deployed on a much wider scale than high school students. Indeed, this was also suggested and encouraged by one of the breakout session groups during the "shark tank" event.

*D. "Shark Tank" Event Results*

The participants in the final ("shark tank") event numbered approximately 55 including the OSU team, 17 high school students, 15 teachers, the mayor, and representatives from local businesses, many of whom also attended the initial event. The data compiled during the "shark tank" event consisted of several surveys summarized below:

- Pre-"shark tank" Survey: 11 questions (Supplement E). This survey was given to teachers and community members and attempted to measure (via a 6-point Likert scale, with 6= "strongly agree") the attendee's opinions on entrepreneurship, technology literacy, their impact on the community, and the attendee's level of personal investment in programs for improving tech literacy and entrepreneurship.
- Post-"shark tank" Survey: the same questions were asked of the same attendees to measure how the project and student work presented impacted their original opinions. These results are discussed in Section F later in this paper.
- Post-project "Then-Now" Survey (Supplement F): This 28-point survey was given to both students and teachers and attempted to measure (via a 5-point Likert scale, 0-4, with 0=strongly disagree, 1=disagree somewhat, 2=not sure, 3=agree somewhat, 4=strongly agree) the impact of the project in 5 categories:
    1. Perceived differences in technology literacy stemming from gender (3 questions)
    2. Self-assessment of technology literacy or entrepreneurial skills (8 questions)
    3. Valuing tech literacy and entrepreneurship for community or self-interests (5 questions)
    4. Levels of interest or enthusiasm for learning and using SCT in general (3 questions)
    5. Levels of ambition to work with SCT for entrepreneurial reasons (5 questions)

The survey statements solicited student respondents to rate their agreement to survey statements *before* the entire PETE-Kit project ("Then") versus their agreement *at the end* of the project ("Now"). This type of retrospective analysis has been shown to elicit larger differences than in traditional pre/post evaluation methods [46]. Additional research has found that such self-assessments are reliable [47] and useful for identifying when learning took place [48]. Teacher respondents did not experience the PETE-Kit project, so their survey compared opinions before the workshop and "shark tank" event (Then) to opinions after both events (Now). Some selected statements used for discussion are shown in Table II.

TABLE II.
SELECT "THEN-NOW" SURVEY QUESTIONS

| Statement Identifier | Statement | Category |
|---|---|---|
| S9 | Girls are better at tech than boys | 1 |
| S15 | Girls and boys are equally good at working with technology applications | 1 |
| S13 | I know what entrepreneurs do for their jobs | 2 |
| S28 | I have strong skills involving SCT | 2 |
| S26 | I have a strong understanding how to turn my idea into a business opportunity | 2 |
| S18 | I am willing to engage in entrepreneurial activity | 2 |
| S19 | I am interested in engaging in entrepreneurial activities that involve tech | 2 |
| S24 | I would like to use SCT to solve personal or community problems | 2 |
| S6 | We learn important things working with PETE kits | 3 |
| S17 | I can understand how PETE Kits can improve tech literacy | 3 |
| S20 | I am aware of how tech can benefit my community | 3 |
| S8 | When we work with tech to solve problems, we use a lot of interesting materials and tools | 4 |
| S25 | I have an interest in pursuing careers related to SCT | 5 |

*E. Then-Now Survey Results*

Table III shows averages of responses across categories and participants. Except for scores in Category 1, the Then and Now scores in the table may be generally interpreted as "agreement" or "disagreement" in the category as a whole. In Category 1, one of the three statements was worded in an opposite sense

> REPLACE THIS LINE WITH YOUR MANUSCRIPT ID NUMBER (DOUBLE-CLICK HERE TO EDIT) <            7(e.g., S9 vs. S15), which tended to neutralize the average scores (2 = neutral).

Category 1 revealed that both student and teachers widely agreed that no gender was "better" than the other with technology, but the teachers showed less variation than students in this respect. For example, S15 teacher average = 3.8/4.0, student average = 3.5. The very low change in this category revealed that the project had no significant impact in shifting this opinion in either participant class. This was an expected result as the curriculum did not include any discussion of gender.

TABLE III.
THEN-NOW SURVEY RESULTS ACROSS 5 CATEGORIES

| Category | Then | Now | Change |
|---|---|---|---|
| 1 (student) | 2.20 | 2.18 | -0.03 |
| 1 (teacher) | 1.53 | 1.51 | -0.02 |
| 2 (student) | 1.81 | 2.90 | 1.09 |
| 2 (teacher) | 2.02 | 2.88 | 0.87 |
| 3 (student) | 2.33 | 3.31 | 0.98 |
| 3 (teacher) | 2.69 | 3.50 | 0.81 |
| 4 (student) | 2.39 | 3.40 | 1.01 |
| 4 (teacher) | 2.36 | 3.07 | 0.71 |
| 5 (student) | 1.98 | 3.03 | 1.05 |
| 5 (teacher) | 2.07 | 3.21 | 1.15 |

Category 2 showed that in assessing participants' self-assessment of tech literacy and entrepreneurial knowledge, both groups showed an increase, with students showing the larger positive change (+1.09) than teachers (+0.87). Some statements had a marked increase in agreement, such as S28, S13, and S26, where the student changes were +1.36, +1.33, and +1.64, respectively. The teacher changes were a more modest +0.9, +0.7, +1.1. We therefore conclude that the students became significantly more confident in their technical skills, but *even more so in entrepreneurial ventures*. This highlights both an elevated importance and effectiveness of entrepreneurship training in our curriculum. On the other hand, teachers became more confident in their tech literacy, but not so much in entrepreneurship. This suggests that the extended time the students had to work with the PETE Kits and their training in how to market a product, over 6 months, versus the teachers' 5-hour professional development workshop, was very beneficial to strengthen participants' perceived technology and entrepreneurial literacy. It also reveals that in our path forward teachers could benefit from closer guidance in adopting and conveying the entrepreneurial portions of the curriculum. This might take the form of extended professional development and dedicated entrepreneurship training for teachers who elect to employ PETE-Kits.

Category 3 survey statements revealed that both students and teachers showed positive gains in understanding the importance and applicability of technology literacy to the community and self. Students started out with neutrality to these statements (2.33) but became agreeable (3.31) for a gain of +0.98. It was also interesting to note that at the start of the program, students were not uniformly convinced that working with the PETE-Kits would help them learn anything "important." But by the end of the program, they indicated that they saw the true value in what the PETE- Kits could bring to their community. Subjectively, this was also clearly demonstrated during the students' "shark tank" presentations to the audience of community members and teachers. Statements S6, S17, and S20 directly addressed this point and saw changes of +1.6, +1.6, and +1.5 respectively, some of the largest positive changes in the survey. It remains to be determined whether this change was due solely to the PETE-Kit and its curriculum or also to the strong guidance provided by the students' advisor/coach. Teachers also began with slight agreement to this category (2.69) but did not increase as much (+0.81) nevertheless becoming quite agreeable (3.51) in the end. Since the teachers did not engage in the PETE-Kit program nor entrepreneurship training, this suggest that their impressions were improved by the PETE-Kit introduction, or perhaps more likely by observing the impact of the program manifested by the "shark tank" presentations by the students.

Category 4 – enthusiasm for learning and using SCT – was similar to categories 2 and 3 in that teachers had substantially weaker changes (+0.71) compared to students (+1.01). Teachers generally began with a rather neutral stance (2.36) to a general interest in learning and using SCT. For a specific example, they scored at an average 2.7 on S8. Conversely, they began rather disagreeable (1.5) to the notion of pursuing a career in SCT (S25). This is not surprising given that teachers already have established careers. However, the noteworthy increase in teacher scores for this category suggests an increased enthusiasm to incorporate PETE-Kit-based training into their own curricula, which again correlates well with subjective observations of the OSU research team during breakout discussions during the teacher workshop and "shark tank" event.

The significantly stronger student response to this category was also revealing. Students also showed neutrality with these statements (2.39) from the start but ended quite agreeable (3.40). Statement S25 (interest in SCT-related career) was notable again in that students disagreed before the project (1.9) and became agreeable (3.0) by the end. We conclude that the PETE-Kit/entrepreneurship project was particularly effective at motivating youth to investigate tech-based careers. The interest shown by the local community at the "shark tank" event was also likely helpful in allowing students to see that this type of work was valued.

Category 5 revealed that both students (+1.05) and teachers (+1.15) grew significantly, and rather equally, in the ambition to use technology specifically for entrepreneurial purposes. Both groups started with neutrality to the statements in this category (students 1.98 and teachers 2.07). Statement S18 was particularly revealing in that teachers grew much more agreeable in willingness to engage in entrepreneurial activities with tech (+1.5). This correlates well with the OSU team's subjective observation during breakout discussions that teachers would be very willing to adopt this curriculum if they had more training. Since no entrepreneurial training was given to the teachers during the workshop, we conclude that their increased enthusiasm stemmed from their observation of the students' entrepreneurial growth exhibited during the "shark tank" presentations. This indirectly validates the effectiveness of the entrepreneurial training of the students during the project.

We also speculate that the student shift in this category may be tied to the students being tasked to solve a community problem. Statements S5 and S24 both addressed using SCT to solve *community* problems and students became significantly more agreeable (+1.2, +1.1 respectively). In addition, in both cases students began as relatively *disagreeable* to the notion (1.8, 1.9). This suggests there was value in the students' minds to using their interests and education in serving a larger purpose (which, again, was reinforced by community response to the "shark tank" event). Moreover, it reveals that by incorporating entrepreneurship, the PETE-Kit curriculum was enabling for this attitude.

Since both teachers and students were positively impacted in this category, it suggests that teachers who incorporate the PETE-Kit curriculum into their existing curriculum could provide rural students with learning opportunities that only students from urban or suburban schools are typically privy to. It remains to be determined more objectively how impactful the student presentations during "shark tank" were to the teachers.

*F. Pre-Post "Shark Tank" Attendee Effectiveness Survey*

The 15 teachers and 16 community members were surveyed with an additional nine questions both immediately before (i.e., as they arrived) and after the "shark tank" event to measure the effectiveness of our approach and the participants' willingness to be part of such work in their own community. This survey was administered on a 6-point Likert scale (1 through 6, 6=strongly agree). Results for select survey questions are demonstrated in Figures 3 and 4 and discussed below.

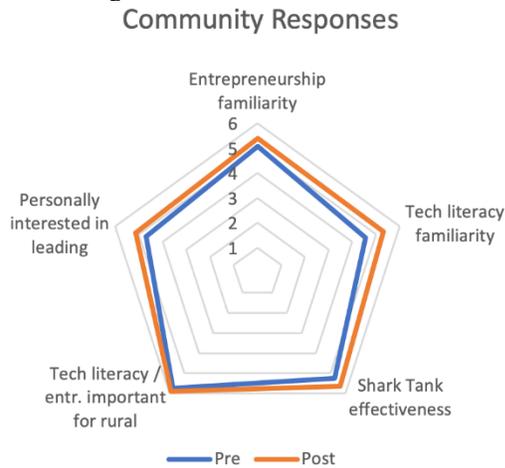

**Fig. 3.** Pre- and post-survey results: Community members

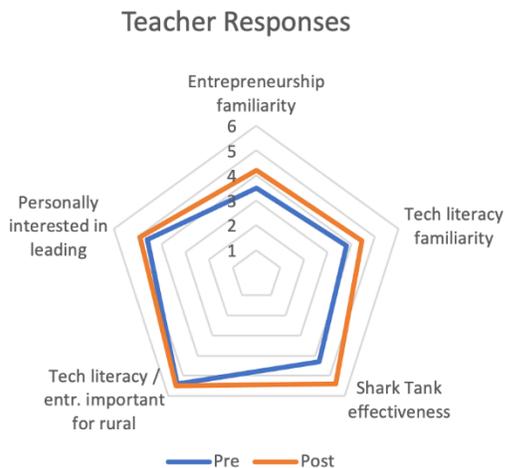

**Fig. 4.** Pre- and post-survey results: Teachers

- "I am familiar with the process of entrepreneurship." Teacher scores increased more (3.50-4.20, Δ = 0.70) than those for community members (5.06-5.38, Δ= 0.32), who were much more confident to begin with. Community attendees included several successful business owners, so this result is not too surprising. The "shark tank" event clearly impacted teachers much more on this topic.
- "I am familiar with tech literacy." Both groups significantly increased their agreement with this statement. In fact, the growth for teachers (3.8-4.45, Δ= 0.75) equaled exactly that for community members (4.56-5.31, Δ=0.75), suggesting that the "shark tank" event was effective at raising awareness of the concept of tech literacy.
- "A competitive approach, such as the "shark tank" event, is an effective way to learn about technology literacy and entrepreneurship." Teachers became much more agreeable to this (4.30-5.40, Δ=1.10) compared to the community (5.25-5.67, Δ= 0.42), who largely agreed with this statement prior to the event. This provides evidence that the competitive element of the "shark tank" did not leave any negative impressions in either group, validating the approach.
- "Technology literacy and entrepreneurship are important factors to revitalize rural economies and improve rural people's quality of life." This statement had high levels of agreement both before and after the event for teachers and community members, suggesting ongoing support for programs of this nature.
- "I am *personally interested* in *leading* projects that improve technology literacy and entrepreneurship in my community." Agreement with this statement saw some increase among both groups but began from relatively high levels. Community interest in personally leading a similar project was over 5.1 at the conclusion of the event, which speaks to the value that local residents saw in the program. This value likely includes impacts from witnessing the project's effects on participating students.

Finally, two questions were posed to both teachers and community members at only the conclusion of the "shark tank" event to measure participant definitions of entrepreneurship and technology literacy. The questions, which allowed more than one selected answer, were:

1) Which of these definitions best matches your idea of entrepreneurship?
    a) ability to work remotely
    b) ability to create new things or services people want
    c) ability to open an independent business
    d) ability to invest in businesses or people
    e) ability to organize and lead teams
2) Which of these definitions best matches your idea of tech literacy?
    a) ability to use a computer and Internet for shopping, reading news, checking weather, basic healthcare
    b) ability to use computer for writing documents, making presentations and spreadsheets
    c) ability to create a webpage on the Internet

   d) ability to use computers, electronics, and the Internet for creative purposes (art, design, invention)
   e) ability to use computers, electronics, and the Internet for business purposes (entrepreneurship, investing, finances)
   f) ability to program a computer to do new things or combine computers, the Internet, and electronics for useful, new applications.

Survey results are shown in Figure 5.

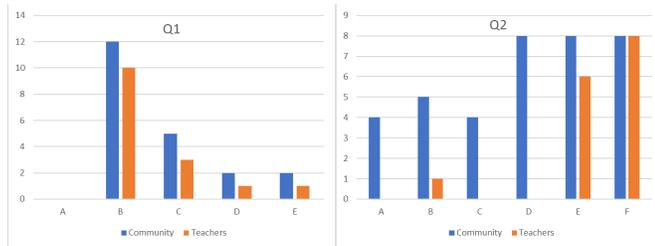

**Fig. 5.** Survey results for definitions of entrepreneurship (Q1) and tech literacy (Q2)

For the definition of entrepreneurship (Q1), both groups overwhelmingly selected option B (ability to create new things or services), with option C (ability to open an independent business) being selected about 1/3 as often. For technology literacy (Q2), community members chose a wide array of definitions while teachers largely concentrated on definitions relating to use in business or new applications.

We conclude that engaging in remote work, leadership, and financing were not perceived as entrepreneurial, whereas creativity was. Future iterations of this curriculum could be tailored to these areas if desired by the local community. The perception of "tech literacy" also highly favored creative purposes but only when those were entrepreneurial in nature. For example, creating a webpage can certainly be creative, but it was not considered representative of tech literacy by the community. It seems that by the end of the "shark tank" event, or perhaps earlier, the survey respondents had a conceptual link between tech literacy and entrepreneurship through the avenue of creativity. These results, combined with those in Figures 3 and 4, suggest that combining entrepreneurship with tech literacy training is a strongly valued combination.

## V. Conclusion

This project resulted in several useful conclusions related to developing a technology-focused entrepreneurial program for rural public schools, which are often restricted with regard to curriculum offerings due to limited teacher numbers and resources. These conclusions are summarized as follows:
1) Rural communities are likely to embrace the idea of helping local students combine technology and entrepreneurship to improve their community.
2) Objective grading of existing technology literacy among rural public high-school students and teachers may demonstrate only a limited understanding (our score for both groups was roughly 38%).
3) The PETE-Kit curriculum demonstrated significant student improvements (i.e., over one full point on a 5-point Likert scale) in retrospective self-assessed technology literacy, valuation of technology / SCTs for community or personal interests, and enthusiasm for / learning SCTs. Teacher improvements were also positive but lower in size.
4) The presence of an embedded and respected "champion" of the project was vital for getting students to buy in and continually work on the project. Our champion was the advisor for the high school robotics team and as such was able to keep regular tabs on student progress and buy-in. More objective measures of this impact are needed in future work.
5) Community involvement is crucial. Engaging the local community at various stages, especially at the beginning and end of the project, not only provides valuable feedback but also fosters a sense of ownership and commitment towards the success of the program.
6) Gaining confidence in entrepreneurship and remote-work is possible. The entrepreneurship training was a critical element of the project and proved to be effective at inspiring confidence in both direct participants (students) and observers (teachers).
7) Hands-on training is effective. The hands-on training approach using SCTs (PETE-kits) proved to be an effective method in enhancing students' understanding and application of technology in real-world scenarios.
8) Ideation events foster creativity. Brainstorming and ideation events serve as effective platforms for students to think creatively, innovate, and come up with entrepreneurial projects that can address local challenges.
9) Digital skills are economic drivers. As rural communities often face limited private-sector job opportunities, equipping students with digital skills can pave the way for local technology-driven entrepreneurship, thereby boosting the local economy.
10) Program scalability and replicability is likely. The success of this program suggests that similar initiatives can be replicated in other rural communities, tailored to their unique challenges and opportunities. High school students need not be the exclusive training target.
11) Long-term impact is promising. While the immediate results of the program are promising, there is potential for long-term impact as students carry forward their skills, aspirations, and entrepreneurial ventures, contributing to sustained community development.
12) Holistic development is effective. Beyond just technical skills, the program also aids in the holistic development of students, enhancing their problem-solving abilities, teamwork, and leadership skills.
13) The convergence of rurality, entrepreneurship, and technology appears to be a very promising combination for effective growth, at least in communities with a strong self-investment.

We conclude by noting that all U.S. states and territories will be receiving significant funds ($2.75 billion) from the federal

government for digital skills training between 2024 and 2028 [6]. One particular group that these funds are required to serve are individuals living in rural locations. The curriculum outlined here represents a timely opportunity for engineering education proponents to apply for and use these funds to engage rural students in applying technology to solve problems in their local communities. Even without direct funding, this type of student-led program can be economically implemented in rural public schools with the expectation that it will be supported by the community and increase familiarity with both technology literacy and local entrepreneurial opportunities.

ACKNOWLEDGMENT

The authors thank Christina Biedny for outstanding graduate student support throughout the duration of the project. They are also thankful to the Rural Renewal Initiative at Oklahoma State University for general support and facilitating outreach to rural communities. Finally, they gratefully thank the Frederick community for being exceptional collaborators and for supporting the local students – and our team – as the project unfolded. All curricular materials noted in this manuscript are available in either the [Supplemental Materials repository](#) and/or upon request from authors.

11
> REPLACE THIS LINE WITH YOUR MANUSCRIPT ID NUMBER (DOUBLE-CLICK HERE TO EDIT) <[34] B. Whitacre, "The diffusion of internet technologies to rural communities: a portrait of broadband supply and demand," *American Behavioral Scientist*, 53,9, pp. 1283-1303, Mar 2010, doi: 10.1177/00027642103616.
[35] A. Van Duersen and J. Van Dijk, "Why Digital Skills are the key to information society," in *Digital skills: Unlocking the information society*, New York, USA: Palgrave Macmillan, 2014, doi: 10.1057/9781137437037.
[36] H. Welser, M. Khan, and M. Dickard, "Digital remediation: social support and online learning communities can help offset rural digital inequality," *Information, Communication and Soc.*, 22,5, pp. 717-723, Apr. 2019, doi: 10.1080/1369118X.2019.1566485.
[37] S. Strover, "Rural internet connectivity," *Telecom. Policy*, 25,5, pp. 331-347, doi: 10.1016/S0308-5961(01)00008-8.
[38] T. Grubesic, "The geodemographic correlates of broadband access and availability in the United States," *Telematics and Informatics*, 21,4, pp. 335-358, Aug 2004, doi: 10.1016/j.tele.2004.02.003.
[39] S. Downes and S. Greenstein, "Universal access and local internet markets in the US," *Research Policy*, 31,7, pp. 1035-1052, Sep 2002, doi: 10.1016/S0048-7333(01)00177-9.
[40] B. Whitacre, R. Gallardo, and S. Strover, "Broadband's contribution to economic growth in rural areas: Moving towards a causal relationship," *Telecom. Policy*, 38,11, pp. 1011-1023, Dec 2014, doi: 10.1016/j.telpol.2014.05.005.
[41] A. Perrin, "Digital gap between rural and nonrural persists," Pew Research Mar 31 2019, available: https://www.pewresearch.org/fact-tank/2019/05/31/digital-gap-between-rural-and-nonrural-america-persists/.
[42] L. Fishbane and A. Tomer, "Broadband adoption is on the rise, but states can do much more," Brookings Institution Sep 10 2019, available: https://www.brookings.edu/blog/the-avenue/2019/10/10/broadband-adoption-is-on-the-rise-but-states-can-do-much-more.
[43] E. Vogels, "Some digital divides persist between rural, urban, and suburban America," Pew Research Aug 19 2021, available: https://www.pewresearch.org/short-reads/2021/08/19/some-digital-divides-persist-between-rural-urban-and-suburban-america/.
[44] D. Moogk, "Minimum viable product and the importance of experimentation in technology startups," *Technology Innovation Mgmt. Review*, 2,3, pp. 23-26, Aug 2012.
[45] National Research Council, *Next Generation Science Standards: For States By States*. Washington, DC: The National Academies Press, 2013. https://doi.org/10.17226/18290.
[46] W. Levinson, G. Gordon, and K. Skeff, "Retrospective versus actual pre-course self-assessments," *Evaluation & the Health Professions*, 13,4, pp. 445-452, Dec 1990, doi: 10.1177/016327879001300406.
[47] N. Brown, D. Dewey, and T. Cox, "Assessing the validity of can-do statements in retrospective (then-now) self-assessment," *Foreign Language Annals*, 47,2, pp. 261-285, Summer 2014, doi: 10.1111/flan.12082.
[48] F. Bhanji, R. Gottesman, W. de Grave, Y. Steinert, and L. Winer, "The retrospective pre-post: A practical method to evaluate learning from an educational program," *Academic Emergency Medicine* 19, 2, pp. 189-194, Feb 2012, doi: 10.1111/j.1553-2712.2011.01270.x.
**Matthew Rutherford** is a Professor and the Johnny Pope Chair and Doctoral Program Coordinator in the School of Entrepreneurship at Oklahoma State University. Previously, Dr. Rutherford was Associate Professor of Management at Virginia Commonwealth University where he oversaw the entrepreneurship program. He has also served on faculty at Gonzaga University. He received his PhD from Auburn University. He has published over 45 peer reviewed articles in top entrepreneurship and management journals and is the author of the book: *Strategic Bootstrapping*. Additionally, he has presented over 100 research manuscripts at international, national, and regional conferences. Dr. Rutherford's experience is in new and small firm consulting. He has provided consulting services to hundreds of organizations of all sizes. His research foci are new venture strategy, family business, and new venture finance.

**Brian Whitacre** is Professor and Jean & Patsy Neustadt Chair in the department of Agricultural Economics at Oklahoma State University. He received his Ph.D. from Virginia Tech. Brian's main area of interest is rural economic development, with a focus on the role that technology can play. He has published over 70 peer-reviewed journal articles, with most exploring the relationship between internet access and rural development. He has developed innovative outreach programs that help small towns benefit from the internet. Brian has won regional and national awards for his research, teaching, and extension programs. Dr. Whitacre also serves as the chair of Oklahoma's Broadband Expansion Council.

**Levi Captain** received his Bachelor of Science degrees in Electrical Engineering and Computer Engineering from Oklahoma State University in Stillwater, OK, USA, in 2022. He is currently pursuing a Master's of Science Degree in Electrical Engineering. Research and areas of interest include terahertz communications, IoT, data Communications, and rural STEM education.

**Sabit Ekin** (SM'21) received his Ph.D. degree in Electrical and Computer Engineering from Texas A&M University, College Station, TX, USA, in 2012. He has four years of industrial experience as a Senior Modem Systems Engineer at Qualcomm Inc., where he received numerous Qualstar awards for his achievements and contributions to the design of cellular modem receivers. He is currently an Associate Professor of Engineering Technology and Electrical & Computer Engineering at Texas A&M University. Prior to this, he was an Associate Professor of Electrical and Computer Engineering at Oklahoma State University. His research interests include the design and analysis of wireless communication and sensing systems, and applications of artificial intelligence and machine learning.

**Julie Angle** received her B.S degree in science education with an emphasis in biology and chemistry from Oklahoma State University (1980), a master's degree in science education from Northwestern Oklahoma State University (1983), and a PhD in science education from Oklahoma State University (2006). Julie has 25 years of high school teaching experience including chemistry, physics, biology, and robotics. In 2008, she was selected as an Albert Einstein Distinguished Educator Fellow and placed with the National Science Foundation in the Office of Cyberinfrastructure. In 2009, she joined the faculty at Oklahoma State University (OSU) as the coordinator for the secondary science education program. She achieved full professorship in 2021, holding the title of "Bill and Billie Dean Buckles Innovation in Teaching Endowed Professor". During her career in education, Julie received multiple awards for teaching excellence and service in education. She served as the coordinator for the Science and Engineering Fair program for the state of Oklahoma and held the office of president for the National Association of Biology Teachers. Julie has hosted hundreds of science teachers in professional development workshops.

**Tom Hensley** serves as the Robotics Instructor and Social Studies Department Chair for Frederick High School. He served in the United States Air Force from January 1988 until February 2014. He served in several combat operations including Operation Provide Promise (Bosnia), Operation Deliberate Force (Bosnia), Operation Deny Flight (Italy), Operation Desert Shield (Saudi Arabia), Operation Enduring Freedom (Qatar), Operation Iraqi Freedom (Qatar/Iraq), and Operation New Dawn (Kyrgyzstan). He has earned multiple awards including 6 Meritorious Service Medals, 5 Air Force Commendation Medals, 3 Air Force Achievement Medals, 1 Joint Meritorious Unit Medal w/gold border, and 6 Meritorious Unit Awards with Valor. His career culminated in his assignment as the Chief of Air Traffic Operations at Altus Air Force Base, ok. He received an Associate's degree in applied Airway Science in 1993, his Bachelor of Science in Social Psychology in 2011 and his Oklahoma Teaching certificate in 2017. He was awarded the Frederick Public Schools Teacher of the Year in 2018, The Frederick Masonic Teacher of today in 2019, and the Professional Oklahoma Educators High School Teacher of the Year for the state of Oklahoma in 2023.

**John F. O'Hara** (SM'19) received his Bachelor of Science degree in electrical engineering from the University of Michigan at Ann Arbor, MI, USA, in 1998, and the Ph.D. degree in electrical engineering from Oklahoma State University, Stillwater, OK, USA, in 2003. He was the Director of Central Intelligence Postdoctoral Fellow with Los Alamos National Laboratory (LANL), Los Alamos, NM, until 2006. From 2006 to 2011, he was with the Center for Integrated Nanotechnologies at LANL and worked on numerous metamaterial projects involving dynamic control over chirality, resonance frequency, polarization, and modulation of terahertz waves. In 2011, he founded a consulting/research company, Wavetech, LLC., Stillwater, OK. In 2017, he joined Oklahoma State University and is now an Associate Professor and the PSO/Albrecht Naeter Professor of ECE. He is a 2023 NSF Early Career grant recipient working in the topic of terahertz wireless communications. He also currently studies IoT, metamaterials, photonic sensing technologies, and rural STEM education. He has 4 patents and around 100 publications in journals and conference proceedings.